\documentclass[twocolumn]{aastex62}

\usepackage{amsmath}
\hypersetup{breaklinks=true,colorlinks=true,citecolor=blue,linkcolor=blue,urlcolor=blue}

\def\hst{\textit{HST}}

\shorttitle{The late-time light curve of SN 2014J}
\shortauthors{Or Graur}

\begin{document}

\title{Late-Time Observations of Type Ia Supernova SN 2014J with the Hubble Space Telescope Wide Field Camera 3}

\correspondingauthor{Or Graur}
\email{or.graur@cfa.harvard.edu}

\author{Or Graur}
\affiliation{Harvard-Smithsonian Center for Astrophysics, 60 Garden St., Cambridge, MA 02138, USA}
\affiliation{Department of Astrophysics, American Museum of Natural History, New York, NY 10024, USA}


\begin{abstract}
 Recent works have studied the late-time light curves of Type Ia supernovae (SNe Ia) when these were older than 500 days past $B$-band maximum light. Of these, SN 2014J, which exploded in the nearby galaxy M82, was studied with the Advanced Camera for Surveys onboard the \textit{Hubble Space Telescope} (\hst) by Yang et al. Here, I report complementary photometry of SN 2014J taken with the \hst\ Wide Field Camera 3 when it was $\sim 360$--$1300$ days old. My \textit{F555W} measurements are consistent with the \textit{F606W} measurements of Yang et al., but the \textit{F438W} measurements are $\sim 1$ mag fainter than their \textit{F475W} measurements. I corroborate their finding that even though SN 2014J has spatially resolved light echoes, its photometry is not contaminated by an unresolved echo. Finally, I compare the \textit{F438W} and \textit{F555W} light curves of SN 2014J to those of the other late-time SNe Ia observed to date and show that more intrinsically luminous SNe have slower light-curve decline rates. This is consistent with the correlation claimed by Graur et al., which was based on a comparison of pseudo-bolometric light curves. By conducting a direct comparison of the late-time light curves in the same filters, I remove any systematic uncertainties introduced by the assumptions that go into constructing the pseudo-bolometric light curves, thus strengthening the Graur et al. claim.
\end{abstract}



\section{Introduction}
\label{sec:intro}

Over the years, several observational techniques have been developed to constrain the nature of the progenitor and explosion mechanism of Type Ia supernovae (SNe Ia; see review by \citealt{2014ARA&A..52..107M}). These methods include, but are not limited to, reconstruction of the delay-time distribution through measurements of SN Ia rates (see \citealt{2017ApJ...848...25M} for the latest summary); analysis of pre-explosion imaging (e.g., \citealt{Li2011fe,2014MNRAS.442L..28G,2014ApJ...790....3K,2018arXiv181104944G}); and multi-wavelength observations (e.g., \citealt{2012ApJ...746...21H,2012ApJ...751..134M,2014ApJ...790...52M,2016ApJ...821..119C}).

Recently, late-time observations of SNe Ia $>500$ days past $B$-band maximum light (hereafter referred to simply as, e.g., $>500$ days) have been suggested as a new way to constrain SN Ia progenitor, explosion, and nebular physics. Several works have shown that, starting at $\sim 800$ days, the decline of the optical light curves of SNe Ia begins to slow down. Until that point, the light curves are dominated by the radioactive decay of $^{56}$Co. The slow-down in the decline rate of the light curves indicates that an additional source of energy is contributing to the heating of the SN ejecta. 

\begin{figure*}
 \centering
 \includegraphics[width=1\textwidth]{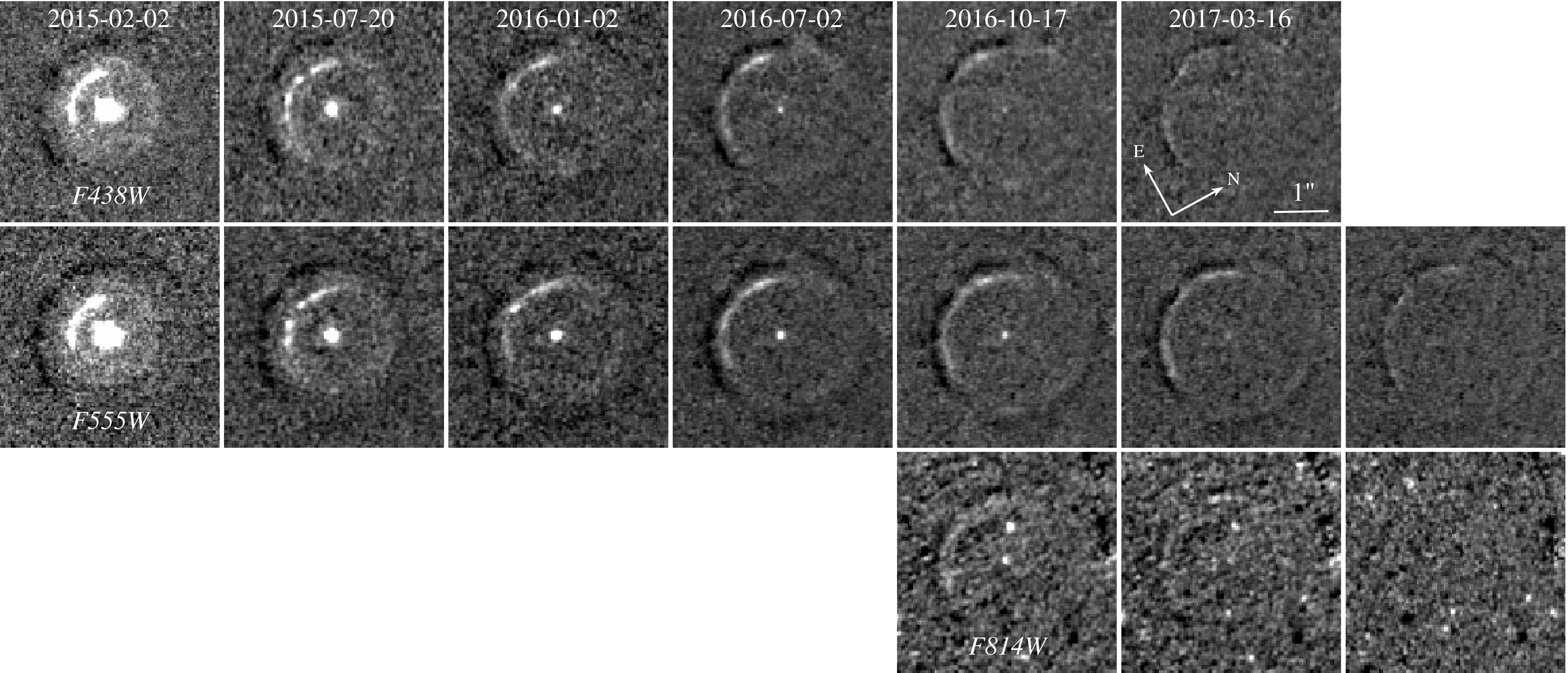}
 \caption{Difference images of SN 2014J in \textit{F438W} (top row), \textit{F555W} (center row), and \textit{F814W} (bottom row). The SN is at the center of each panel, which is $4^{\prime\prime}$ on a side. The spatially resolved light echoes (white arc and diffuse ``ring'') can be seen advancing with time. The dark arc marks the location of the light echo in the template image used in the subtraction. The SN is no longer detected in the final difference image in each filter.}
 \label{fig:diffs}
\end{figure*}

\citet{2016ApJ...819...31G} showed that the light curve of SN 2012cg was consistent with a combination of $^{56}$Co and $^{57}$Co radioactive decays (as predicted by \citealt{2009MNRAS.400..531S}). The same was shown by \citet{2017ApJ...841...48S} for SN 2011fe and by \citet{2018ApJ...852...89Y} for SN 2014J. According to \citet{2017ApJ...841...48S}, the late-time photometry of SN 2011fe was precise enough to rule out the single-degenerate progenitor scenario \citep{Whelan1973}, but \citet{2017MNRAS.468.3798D} and \citet{2017MNRAS.472.2534K} showed that similar data could also be fit with models that assumed either atomic freeze-out \citep{1993ApJ...408L..25F,2015ApJ...814L...2F} or a changing fraction of positron trapping (which could be due, for example, to temporal variations in the magnetic field of the ejecta; \citealt{2014ApJ...795...84P}).
  
\citet{2018ApJ...859...79G,2018ApJ...866...10G} analyzed the late-time light curves of five SNe Ia and claimed a possible correlation between the decline rate of the late-time light curve and the intrinsic luminosity of the SNe. This correlation was challenged by \citet{2018ApJ...857...88J}; based on a single epoch of photometric measurements of SN 2013aa at $\approx 1500$ days, they claimed that it significantly fell of the correlation.
  
The correlation claimed by \citet{2018ApJ...859...79G,2018ApJ...866...10G} suffers from two main sources of uncertainty: a statistical uncertainty due to the small number of SNe observed at $>500$ days and a systematic uncertainty due to the different assumptions that go into the construction of the pseudo-bolometric light curves on which the correlation is based. Here, I attempt to remove the second source of uncertainty by directly comparing observations of the SNe studied by \citet{2018ApJ...859...79G,2018ApJ...866...10G} in the same filters. This is made possible by new observations of SN 2014J.

The late-time light curve of SN 2014J was studied by \citet{2018ApJ...852...89Y} using data taken with the Advanced Camera for Surveys (ACS) onboard the \textit{Hubble Space Telescope} (\hst). In Section~\ref{sec:data}, I present complementary data taken with the \hst\ Wide Field Camera 3 (WFC3). I show that the new photometry is consistent with that of \citet{2018ApJ...852...89Y} but, crucially, includes data in \textit{F438W} and \textit{F555W}, which allows a direct comparison with similar photometry of SNe 2011fe, 2012cg, 2015F, and ASASSN-14lp. In Section~\ref{subsec:le}, I confirm the conclusion reached by \citet{2018ApJ...852...89Y} that, although SN 2014J has spatially resolved light echoes, the photometry of the SN itself is not contaminated by an unresolved echo. In Section~\ref{subsec:correlation}, I show that the correlation claimed by \citet{2018ApJ...859...79G,2018ApJ...866...10G} is still apparent when examining the observed---instead of pseudo-bolometric---light curves. I summarize my findings in Section~\ref{sec:discuss}.


\section{Observations and Photometry}
\label{sec:data}

SN 2014J was discovered by \citet{2014CBET.3792....1F} on 2014 Jan. $21.8$ in the nearby ($3.3$ Mpc; \citealt{2014MNRAS.443.2887F}) starburst galaxy M82. With a SiFTO stretch value of $1.086 \pm 0.010$ \citep{2008ApJ...681..482C,2015MNRAS.454.3816C}, SN 2014J is a luminous but still normal SN Ia. According to \citet{2015ApJ...798...39M}, the SN reached peak $B$-band light on 2014 Feb. $1.7$ (MJD $56689.74 \pm 0.13$); all phases in this work are measured relative to this date.  

For this work, I used WFC3 images of SN 2014J in the broad-band filters \textit{F438W}, \textit{F555W}, and \textit{F814W} taken under \hst\ programs GO--13626 (PI: Lawrence), GO--14146 (PI: Lawrence), GO--14700 (PI: Sugerman), and SNAP--15166 (PI: Filippenko). These programs imaged SN 2014J between 2014 Mar. 11 and 2017 Dec. 17. In this work I concentrate on the late-time light curve of SN 2014J, and so only present data starting at 2015 Feb. 2, when the SN was $\sim 365$ days old. 

\begin{deluxetable}{LCcCC}
 \tablecaption{Observation log for SN 2014J. \label{table:mags_14J}}
 \tablehead{
 \colhead{MJD}    & \colhead{Phase}  & \colhead{Filter} & \colhead{Exposure} & \colhead{Magnitude} \\
 \colhead{(days)} & \colhead{(days)} & \colhead{}       & \colhead{(s)}      & \colhead{(Vega mag)} 
 }
 \decimals
 \startdata
  57055.2 & 365.5  & {\it F555W} & 384  & 18.824 \pm 0.002 \\
  57055.2 & 365.5  & {\it F438W} & 1536 & 19.962 \pm 0.003 \\
  57223.1 & 533.4  & {\it F555W} & 432  & 21.36  \pm 0.01  \\
  57223.1 & 533.4  & {\it F438W} & 1344 & 22.32  \pm 0.01  \\
  57389.2 & 699.5  & {\it F555W} & 432  & 22.36  \pm 0.03  \\
  57389.3 & 699.6  & {\it F438W} & 1344 & 23.73  \pm 0.05  \\
  57571.1 & 881.4  & {\it F555W} & 1720 & 23.57  \pm 0.06  \\
  57571.2 & 881.5  & {\it F438W} & 2420 & 24.86  \pm 0.09  \\
  57678.1 & 988.4  & {\it F555W} & 1664 & 24.13  \pm 0.10  \\
  57678.1 & 988.4  & {\it F814W} & 1640 & 23.48  \pm 0.19  \\
  57678.1 & 988.4  & {\it F438W} & 2600 & 25.67  \pm 0.18  \\
  57828.0 & 1138.3 & {\it F555W} & 1664 & 24.81  \pm 0.19  \\
  57828.0 & 1138.3 & {\it F814W} & 1640 & >24.4            \\
  57828.0 & 1138.3 & {\it F438W} & 2600 & >26.1            \\
  57985.7 & 1296.0 & {\it F555W} & 1664 & >25.3            \\
  57985.7 & 1296.0 & {\it F814W} & 1640 & >24.4            \\
 \enddata
 \textbf{Note.} All photometry is measured using aperture photometry with a $0^{\prime\prime}.4$-diameter aperture. Upper limits represent a point source with a S/N ratio of 3 at the location of the SN.
\end{deluxetable}

I used the {\sc AstroDrizzle} task included in the {\sc DrizzlePac} Python package\footnote{\url{http://drizzlepac.stsci.edu/}} \citep{2012AAS...22013515H} to align the \hst\ {\sc FLC} images and remove cosmic rays and bad pixels. Next, I aligned the images using the IRAF\footnote{\url{http://iraf.noao.edu/}} routines {\sc xregister} and {\sc wcscopy} \citep{1986SPIE..627..733T,1993ASPC...52..173T}. Finally, I created difference images of the SN by subtracting the last epoch in each filter from all earlier epochs. The resultant difference images are shown in Figure~\ref{fig:diffs}. A visual inspection does not detect the SN in the template images or in the final difference image in each filter.

\begin{figure}
 \centering
 \includegraphics[width=0.47\textwidth]{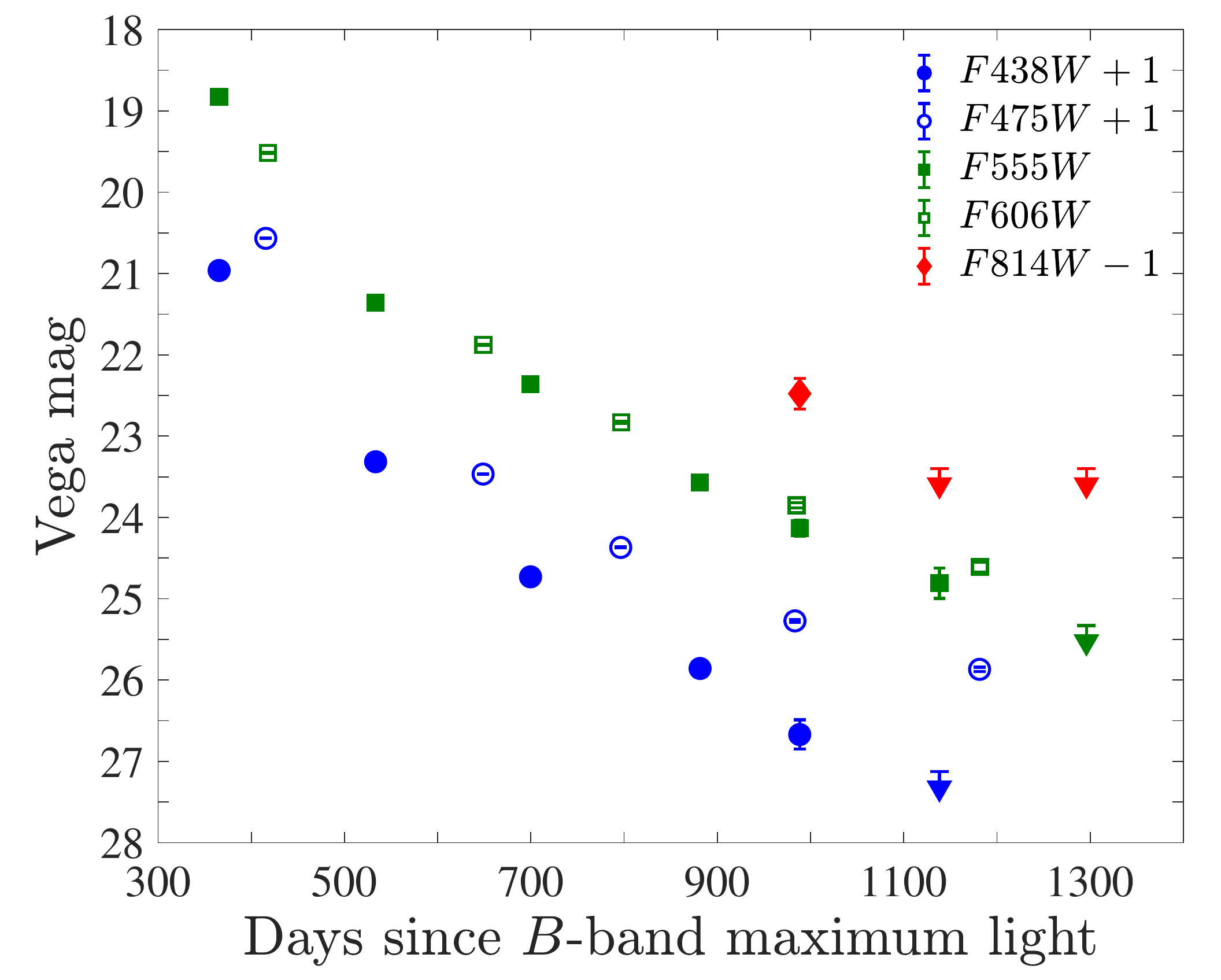}
 \caption{Aperture photometry of SN 2014J in \textit{F438W} (blue filled circles), \textit{F555W} (green filled squares), and \textit{F814W} (red diamonds), compared to the \citet{2018ApJ...852...89Y} photometry in \textit{F475W} (blue open circles) and \textit{F606W} (green open squares). For clarity, the \textit{F438W}, \textit{F475W}, and \textit{F814W} measurements have been offset by $+1$, $+1$, and $-1$ mags, respectively. Downturned arrows represent $3\sigma$ upper limits on the magnitude of the SN in images where it is no longer detected. As I explain in Section~\ref{sec:data}, I attribute the $\approx 1$ mag difference between the \textit{F438W} and \textit{F475W} measurements to the shorter spectral coverage of \textit{F438W}. The \citet{2018ApJ...852...89Y} photometry was converted from AB to Vega mags using the respective \hst\ ACS zero points.}
 \label{fig:phot}
\end{figure}

SN 2014J is known to have spatially resolved light echoes \citep{2015ApJ...804L..37C,2017ApJ...834...60Y}, which are apparent in the difference images in Figure~\ref{fig:diffs}. To ensure that the light echoes did not contaminate the photometry of the SN, I performed aperture photometry with a $0^{\prime\prime}.4$-diameter aperture, and used a sky annulus with a radius larger than the farthest reach of the resolved light echoes. I applied the latest aperture corrections, as described by \citet{2016wfc..rept....3D}.\footnote{\url{http://www.stsci.edu/hst/wfc3/phot_zp_lbn}} The resultant photometry, in Vega mags, is presented in Table~\ref{table:mags_14J} and Figure~\ref{fig:phot}. In epochs where the SN is undetected, I measured $3\sigma$ upper limits by estimating the magnitude at which the SN would have a signal-to-noise (S/N) ratio of 3.

Unlike the clean \textit{F438W} and \textit{F555W} difference images, the \textit{F814W} difference images shown in Figure~\ref{fig:diffs} are speckled with other point sources aside from SN 2014J. A close look at some of these point sources shows that they appear, with different brightnesses, in multiple epochs. Thus, these point sources are likely to be variable stars that, while visible in \textit{F814W}, are extincted in \textit{F438W} and \textit{F555W}.

The location of SN 2014J is known to suffer from high host-galaxy extinction. \citet{2015ApJ...798...39M}, for example, measure a host-galaxy $E(B-V) = 1.23 \pm 0.06$ mag and $R_V = 1.46$. Based on these values, and assuming a \citet{1989ApJ...345..245C} extinction law, throughout this work I use host-galaxy extinction values of $A_{\it F438W} = 3.2$, $A_{\it F555W} = 1.9$, and $A_{\it F814W} = 0.7$ mags. These values are consistent with the $B$, $V$, and $I$ extinctions measured by \citet{2014MNRAS.443.2887F} from the early-phase light curve of SN 2014J. For line-of-sight Galactic extinction, I rely on the \citet{2011ApJ...737..103S} measurements of $A_{\it F438W} = 0.6$, $A_{\it F555W} = 0.5$, and $A_{\it F814W} = 0.2$ mags.

As shown in Figure~\ref{fig:phot}, the \textit{F555W} photometry measured here is consistent with the \textit{F606W} photometry measured by \citet{2018ApJ...852...89Y}. This is expected, as both filters cover roughly the same spectral range. On the other hand, the \textit{F438W} photometry is systematically fainter by $\approx 1$ mag, which is probably due to the fact that the ACS \textit{F475W} filter is roughly $2.4$ times wider than the WFC3 \textit{F438W} filter and so encompasses a prominent spectral feature at $\sim 5400$~\AA\ (likely a combination of [Fe II] and [Fe III]; \citealt{2015MNRAS.454.1948G}) that is left out of the \textit{F438W} filter.

\section{Analysis}
\label{sec:analysis}

In this section, I use the photometry measured in Section~\ref{sec:data} to show that SN 2014J is not contaminated by an unresolved light echo (Section~\ref{subsec:le}) and that an analysis of the late-time \textit{F438W} and \textit{F555W} light curves of SNe Ia is still consistent with the \citet{2018ApJ...859...79G,2018ApJ...866...10G} claim that more luminous SNe Ia have flatter light curves at late times (Section~\ref{subsec:correlation}).

\subsection{Light echoes}
\label{subsec:le}

To date, several SNe Ia have been shown to exhibit light echoes, which are produced when the light of the SN is reflected off nearby dust sheets. Unresolved light echoes will contaminate the photometry of the SN and cause its light curve to flatten out at late epochs \citep{1994ApJ...434L..19S,1999ApJ...523..585S,2006ApJ...652..512Q,2001ApJ...549L.215C,2008ApJ...677.1060W,2015ApJ...805...71D,2018arXiv181011936W}. 

\begin{figure}[t!]
 \centering
 \includegraphics[width=0.47\textwidth]{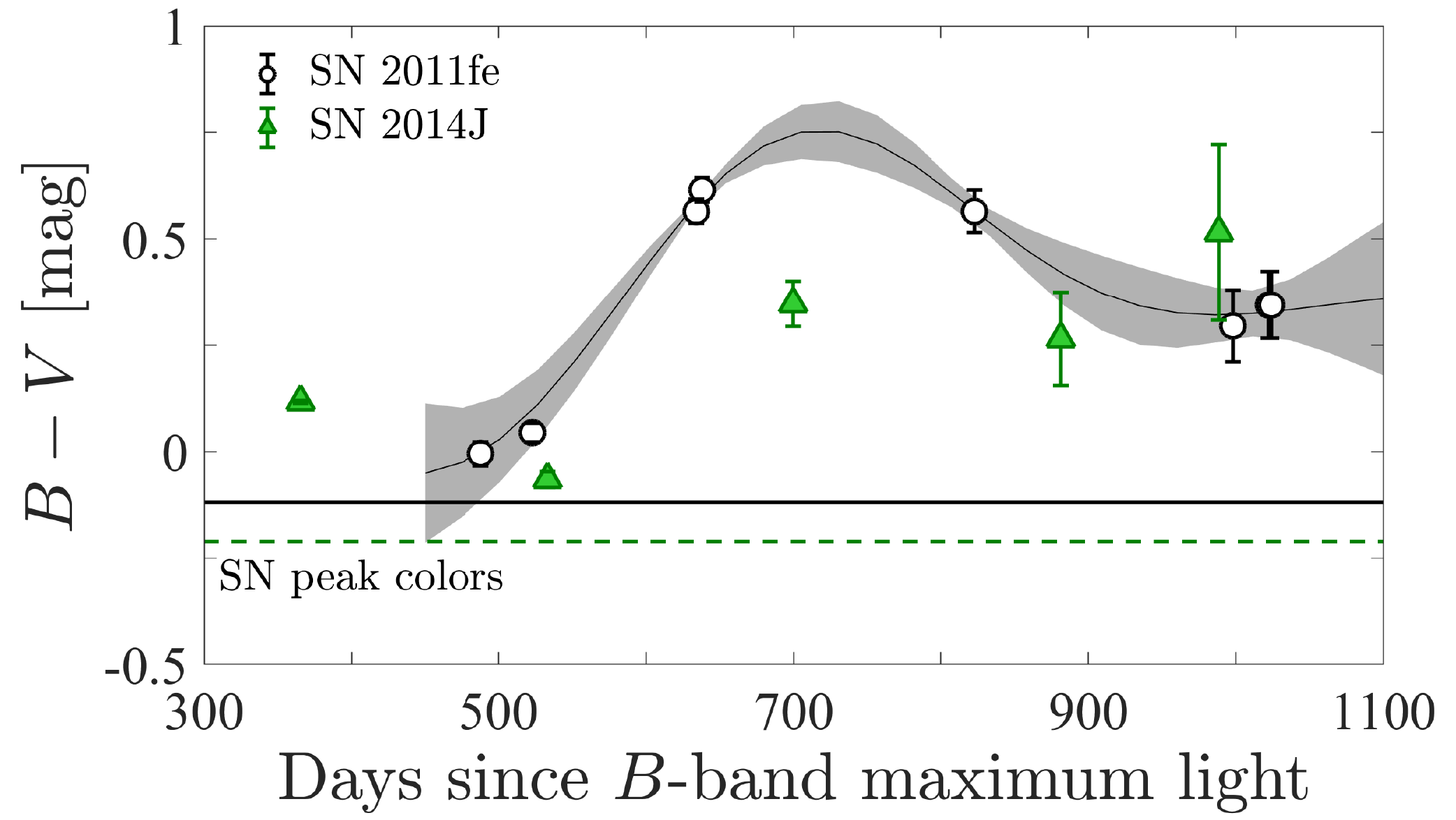}
 \caption{A comparison between the $B-V$ colors of SN 2014J (green triangles) and SN 2011fe (black circles). The dashed and solid curves represent the $B-V$ colors of the two SNe, respectively, at $B$-band maximum light. The gray shaded band connecting the colors of SN 2011fe is a Gaussian Process regression; the width of the band represents the 68\% uncertainty of the fit. The $B-V$ colors of SN 2014J are offset from those of SN 2011fe but show the same trend, i.e., they are redder than the SN was at peak. A similar comparison of a $V-I$ color at $988.4$ days shows that both SNe are bluer in that color than at peak (see Section~\ref{subsec:le}). Taken together, these data rule out an unresolved light echo.}
 \label{fig:color}
\end{figure}

To test whether the photometry of SN 2014J is contaminated by an unresolved light echo, I repeat the exercise done by \citet{2018ApJ...859...79G,2018ApJ...866...10G} and compare, in Figure~\ref{fig:color}, the late-time $B-V$ colors of SN 2014J to those of SN 2011fe \citep{2017ApJ...841...48S}, which is known to be free of light echoes \citep{2015MNRAS.454.1948G,2017ApJ...841...48S}. I also compare these colors to those of the two SNe at peak (as measured by \citealt{2013NewA...20...30M} and \citealt{2014MNRAS.443.2887F} for SNe 2011fe and 2014J, respectively). All magnitudes have been converted from the WFC3 filter sets to the Johnson-Cousins $BVI$ filters and corrected for Galactic and host-galaxy extinctions. The filter-set conversions were calculated by comparing synthetic photometry of a late-time spectrum of SN 2011fe at 981 days (see \citealt{2015MNRAS.454.1948G,2018ApJ...859...79G}).

The $B-V$ colors of both SNe are redder at late times than they were at peak. SN 2014J also has a $V-I$ color of $-1.2 \pm 0.2$ mag at $988.4$ days, which is bluer than the peak $V-I$ color of $\sim -0.3$ mag. This behavior, where the SN is bluer in $V-I$ but redder in $B-V$, is consistent with the late-time colors of SNe 2011fe, 2015F, and ASASSN-14lp \citep{2017ApJ...841...48S,2018ApJ...859...79G,2018ApJ...866...10G}. This is inconsistent with a light echo, which would scatter the SN light to bluer wavelengths and make it appear bluer than it was at peak in all colors. Thus, even though SN 2014J has spatially resolved light echoes, it appears that the aperture photometry conducted here is not contaminated by an unresolved echo, consistent with the findings of \citet{2018ApJ...852...89Y}.

\subsection{The decline rate of the late-time light curve}
\label{subsec:correlation}

\begin{figure*}
 \centering
 \begin{tabular}{cc}
  \includegraphics[width=0.47\textwidth]{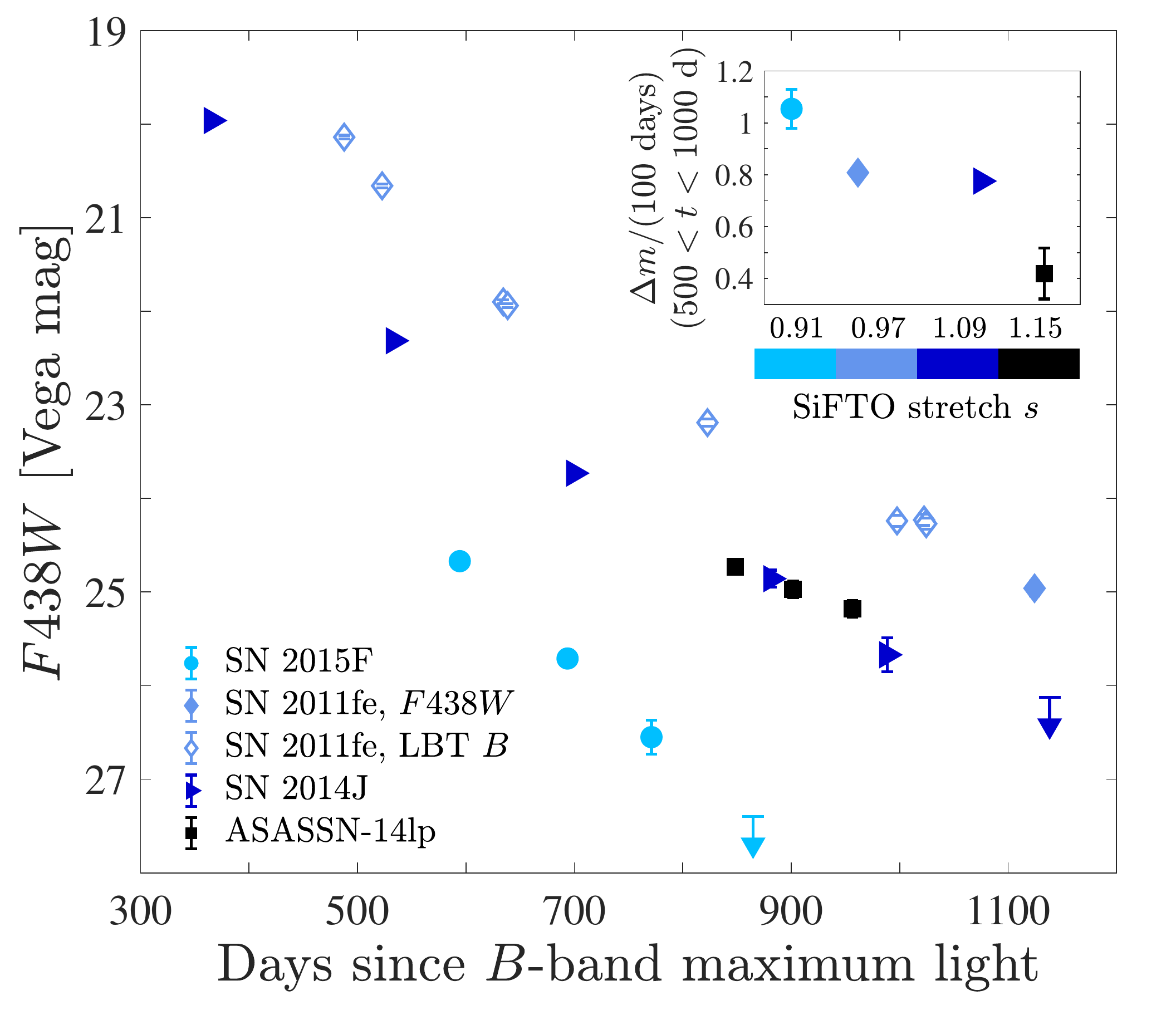} &
  \includegraphics[width=0.47\textwidth]{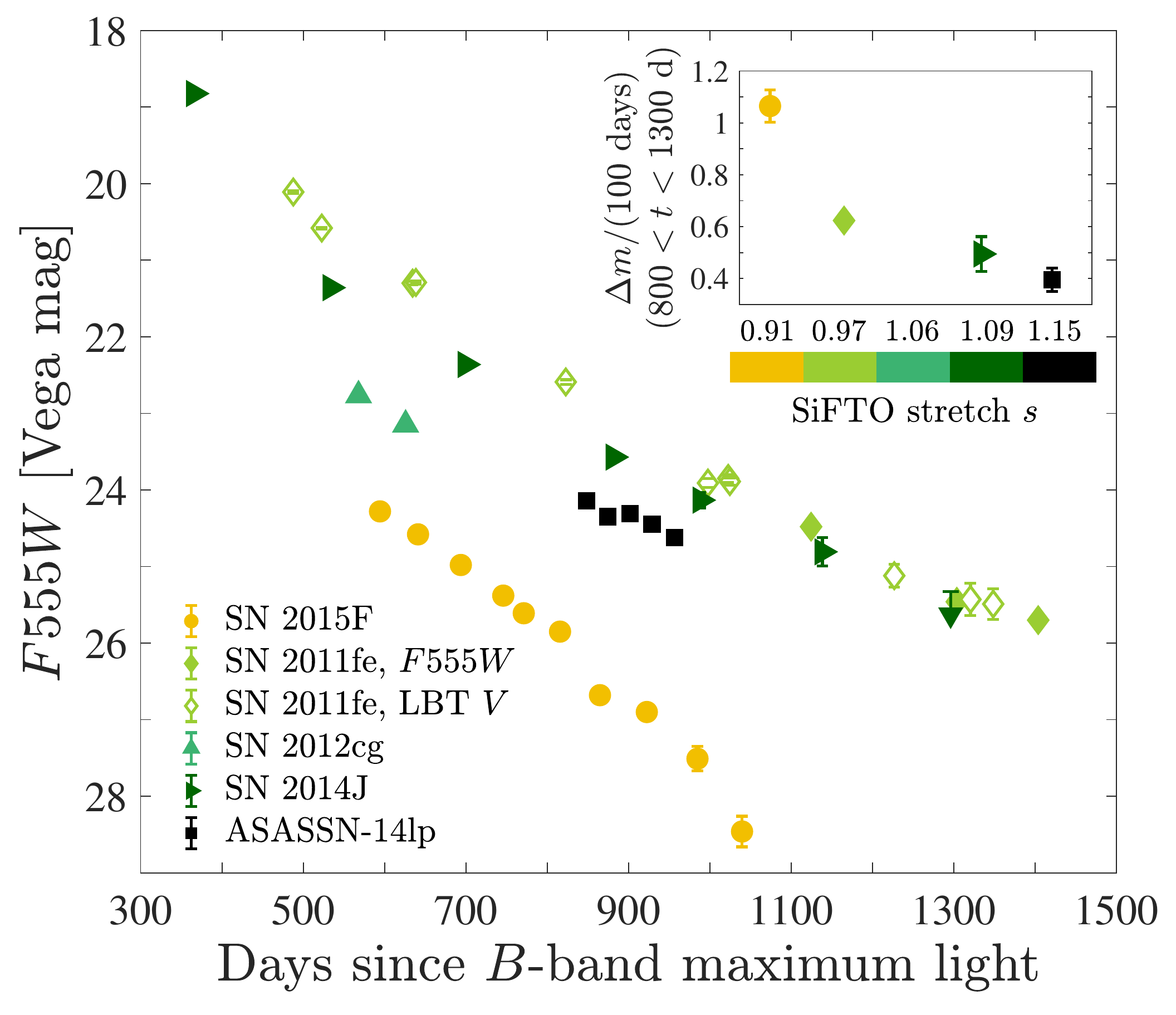} \\
 \end{tabular}
 \caption{The \textit{F438W} (left panel) and \textit{F555W} (right panel) light curves of SN 2014J (right-facing arrows) compared to those of SNe observed in the same filters. SN 2011fe is represented by both \hst\ measurements (filled diamonds) and LBT $B$- and $V$-band measurements (open diamonds). The colormap in each panel is chosen to show that more luminous SNe Ia (with higher stretch values), marked by darker hues, have light curves that decline more slowly at late times. The insets in both panels show the decline rate of the light curve based on a linear fit to the data at $500<t<1000$ days for \textit{F438W} and $800<t<1300$ days (when the light curve is expected to begin to deviate from pure $^{56}$Co decay) for \textit{F555W}.}
 \label{fig:lums}
\end{figure*}

Previous papers on the late-time light curves of SNe Ia used multi-band observations to construct a ``pseudo-bolometric'' light curve in the optical wavelength range of $\sim 3500$--$10000$~\AA. Briefly, in each epoch, a spectrum of the SN is morphed to fit the observed photometry, which is first corrected for Galactic and host-galaxy extinction. A pseudo-bolometric flux is then measured by integrating the morphed spectrum over the observed wavelength range. Finally, the pseudo-bolometric flux is converted to a luminosity, taking into account the distance measured to either the SN or its host galaxy. This technique, though instructive, might introduce systematic uncertainties for several reasons:
\begin{enumerate}
 \item Only SN 2011fe has spectra taken 500--1000 days after explosion \citep{2015MNRAS.454.1948G,2015MNRAS.448L..48T}. The pseudo-bolometric light curves of SNe 2015F, ASASSN-14lp, and 2014J were also constructed using these spectra \citep{2018ApJ...859...79G,2018ApJ...866...10G,2018ApJ...852...89Y}. 
 \item Those spectra only sample SN 2011fe at 593, 981, and 1034 days. According to \citet{2015MNRAS.454.1948G}, SN 2011fe shows little spectral evolution between the 593- and 981-day spectra. It is tacitly assumed that this is true of other SNe Ia as well.
 \item Some SNe did not have full multi-band coverage throughout the phase range in which the pseudo-bolometric light curve was constructed, and extrapolations were used to fill in the gaps (e.g., SN 2015F and ASASSN-14lp).
 \item Although Galactic line-of-sight extinction towards the SNe is well measured, there is still an ongoing debate as to the extinction law (parameterized by the total-to-selective extinction ratio, $R_V$) at the location of SNe Ia (e.g., \citealt{2010AJ....139..120F,Chotard2011,2014ApJ...780...37S,2015MNRAS.453.3300A}).
 \item The measurement uncertainties of the distances to the SNe are not propagated into the final luminosities. This would impact the scaling of the resultant pseudo-bolometric light curve, which could bring into question the interesting claim that the late-time light curves first converge at $\sim 600$ days before diverging at different decline rates \citep{2018ApJ...866...10G,2018ApJ...857...88J}.
\end{enumerate}

\floattable
\begin{deluxetable}{lCCCCCCC}
 \tablecaption{Light curve decline rates. \label{table:declines}}
 \tablehead{
 \colhead{Supernova} & \colhead{Stretch} & \colhead{$\Delta$\textit{F438W}} & \colhead{$\chi2/{\rm DOF}$} & \colhead{$\Delta$\textit{F555W}} & \colhead{$\chi2/{\rm DOF}$} & \colhead{$\Delta$\textit{F555W}} & \colhead{$\chi2/{\rm DOF}$} \\
 \colhead{} & \colhead{} & \colhead{mag/(100 days)} & \colhead{} & \colhead{mag/(100 days)} & \colhead{} & \colhead{mag/(100 days)} & \colhead{} \\
 \colhead{} & \colhead{} & \colhead{$500<t<1000~{\rm days}$} & \colhead{} & \colhead{$500<t<1000~{\rm days}$} & \colhead{} & \colhead{$800<t<1300~{\rm days}$} 
 }  
 \decimals
 \startdata
  SN 2015F    & 0.906 \pm 0.005 & 1.05 \pm 0.07 & 0.02/1 & 0.77 \pm 0.02 & 26/7 & 1.07 \pm 0.06 & 11/3  \\
  SN 2011fe   & 0.969 \pm 0.010 & 0.81 \pm 0.01 & 260/3  & 0.67 \pm 0.01 & 17/3 & 0.62 \pm 0.01 & 13/4  \\
  SN 2012cg   & 1.063 \pm 0.011 & \cdots        & \cdots & 0.66 \pm 0.03 & \cdots & \cdots & \cdots \\
  SN 2014J    & 1.086 \pm 0.010 & 0.76 \pm 0.02 & 11/2   & 0.62 \pm 0.01 & 2/2  & 0.49 \pm 0.07 & 0.1/1 \\
  ASASSN-14lp & 1.150 \pm 0.050 & 0.42 \pm 0.10 & 0.03/1 & 0.39 \pm 0.05 & 8/3  & 0.39 \pm 0.05 & 8/3   \\
 \enddata
\end{deluxetable}

Here, I remove these systematic uncertainties by foregoing the construction of a pseudo-bolometric light curve for SN 2014J. Instead, in Figure~\ref{fig:lums} I compare the \textit{F438W} and \textit{F555W} photometry of SN 2014J measured in Section~\ref{sec:data} to those of SNe 2011fe, 2012cg, 2015F, and ASASSN-14lp in the same filters. The \textit{F555W} comparison is of most value, since previous works have shown that SNe Ia are most luminous in this filter. Moreover, there is some evidence that the \textit{F555W} flux should trace the bolometric light curve. For example, \citet{2001ApJ...559.1019M} showed that between 50 and 600 days, a constant fraction of the luminosity of SNe Ia is emitted in the $V$ band, which is comparable to \textit{F555W}.

The photometry of the different SNe in Figure~\ref{fig:lums} is not corrected for Galactic and host-galaxy extinction. Such corrections, with their accompanying systematic uncertainties, would only impact the scaling of the measurements, which is not important for the analysis below.

As a simple test of the \citet{2018ApJ...859...79G,2018ApJ...866...10G} correlation, I perform a linear fit to the photometry of each SN to measure the decline rates of their light curves. The resultant rates are presented in Table~\ref{table:declines} and shown in the insets of Figure~\ref{fig:lums} as a function of the stretch, $s$, of the SNe. The latter is correlated with the intrinsic luminosity of the SNe through the \citet{1993ApJ...413L.105P} relation. 

Ideally, the decline rate should be measured at $t>800$ days, when previous works have shown that the late-time light curve begins to appreciably deviate from domination by pure $^{56}$Co radioactive decay. This is possible in \textit{F555W}, but not in \textit{F438W}, where there are fewer measurements. For the latter filter, I measure the decline rate at $500<t<1000$ days. The decline rates in \textit{F555W} are measured in two phase ranges: $800<t<1300$ days, shown in Figure~\ref{fig:lums}, and $500<t<1000$ days, to facilitate a comparison with \textit{F438W}. 

SN 2011fe only has \hst\ observations starting at $\approx 1100$ days, but \citet{2017ApJ...841...48S} also provide earlier photometry in $B$ and $V$. Because the latter measurements are consistent with the \hst\ measurements taken at similar times (as shown in Figure~\ref{fig:lums}), I include them in my analysis. For SN 2012cg, which only has two \textit{F555W} measurements, I measure the decline rate between those two visits and use it as an estimate of the decline rate at $500<t<1000$ days.

The light curve decline rates in both filters show the same trend as the pseudo-bolometric light curves in previous works, i.e., more luminous SNe Ia (with larger stretch values) exhibit flatter late-time light curves. However, as also noted by previous works, a larger SN sample is required to substantiate this trend.


\section{Conclusions}
\label{sec:discuss}

I have presented new \hst\ photometry of SN 2014J when it was $\sim 360$--$1300$ days old. These observations, taken with WFC3, complement the \hst\ ACS observations of this SN presented by \citet{2018ApJ...852...89Y}. 

I have used these data to verify that, although SN 2014J exhibits spatially resolved light echoes, the photometry of the SN itself is not contaminated by an unresolved echo. 

Whereas previous works used the photometry of each SN to construct a pseudo-bolometric light curve, in this work I conducted a direct comparison of the \textit{F438W} and \textit{F555W} light curves of SNe 2011fe, 2012cg, 2014J, 2015F, and ASASSN-14lp. This removes any systematic uncertainties introduced by the construction of the pseudo-bolometric light curve, as well as the use of extinction corrections and distance measurements. 

I have measured the light curve decline rates of the five SNe Ia mentioned above at $500<t<1300$ days in both \textit{F438W} and \textit{F555W} and have shown that these are consistent with the correlation claimed by \citet{2018ApJ...859...79G,2018ApJ...866...10G}: more luminous SNe Ia (with higher stretch values) have slower-declining late-time light curves. However, although in this work I have minimized the sources of systematic uncertainty that might affect this claim, the main source of uncertainty---the small sample size---remains unchanged. A larger sample of late-time SNe Ia, observed with the same filters analyzed here, is required to definitively test the \citet{2018ApJ...859...79G,2018ApJ...866...10G} correlation.


\section*{Acknowledgments}

I thank the anonymous referee for helpful comments and suggestions. O.G. was supported by NASA through \hst-GO--14611 and \hst-GO-15415. This work is based on data obtained with the NASA/ESA {\it Hubble Space Telescope}, all of which was obtained from MAST. Support for MAST for non-\hst\ data is provided by the NASA Office of Space Science via grant NNX09AF08G and by other grants and contracts. This research has made use of NASA's Astrophysics Data System and the NASA/IPAC Extragalactic Database (NED) which is operated by the Jet Propulsion Laboratory, California Institute of Technology, under contract with NASA. 


\software{DrizzlePac \citep{2012AAS...22013515H}, IRAF \citep{1986SPIE..627..733T,1993ASPC...52..173T}}



\begin{thebibliography}{}
\expandafter\ifx\csname natexlab\endcsname\relax\def\natexlab#1{#1}\fi

\bibitem[{{Amanullah} {et~al.}(2015){Amanullah}, {Johansson}, {Goobar},
  {Ferretti}, {Papadogiannakis}, {Petrushevska}, {Brown}, {Cao}, {Contreras},
  {Dahle}, {Elias-Rosa}, {Fynbo}, {Gorosabel}, {Guaita}, {Hangard}, {Howell},
  {Hsiao}, {Kankare}, {Kasliwal}, {Leloudas}, {Lundqvist}, {Mattila}, {Nugent},
  {Phillips}, {Sandberg}, {Stanishev}, {Sullivan}, {Taddia}, {{\"O}stlin},
  {Asadi}, {Herrero-Illana}, {Jensen}, {Karhunen}, {Lazarevic}, {Varenius},
  {Santos}, {Sridhar}, {Wallstr{\"o}m}, \& {Wiegert}}]{2015MNRAS.453.3300A}
{Amanullah}, R., {Johansson}, J., {Goobar}, A., {et~al.} 2015, \mnras, 453,
  3300

\bibitem[{{Cappellaro} {et~al.}(2001){Cappellaro}, {Patat}, {Mazzali},
  {Benetti}, {Danziger}, {Pastorello}, {Rizzi}, {Salvo}, \&
  {Turatto}}]{2001ApJ...549L.215C}
{Cappellaro}, E., {Patat}, F., {Mazzali}, P.~A., {et~al.} 2001, \apjl, 549,
  L215

\bibitem[{{Cardelli} {et~al.}(1989){Cardelli}, {Clayton}, \&
  {Mathis}}]{1989ApJ...345..245C}
{Cardelli}, J.~A., {Clayton}, G.~C., \& {Mathis}, J.~S. 1989, \apj, 345, 245

\bibitem[{{Childress} {et~al.}(2015){Childress}, {Hillier}, {Seitenzahl},
  {Sullivan}, {Maguire}, {Taubenberger}, {Scalzo}, {Ruiter}, {Blagorodnova},
  {Camacho}, {Castillo}, {Elias-Rosa}, {Fraser}, {Gal-Yam}, {Graham}, {Howell},
  {Inserra}, {Jha}, {Kumar}, {Mazzali}, {McCully}, {Morales-Garoffolo},
  {Pandya}, {Polshaw}, {Schmidt}, {Smartt}, {Smith}, {Sollerman}, {Spyromilio},
  {Tucker}, {Valenti}, {Walton}, {Wolf}, {Yaron}, {Young}, {Yuan}, \&
  {Zhang}}]{2015MNRAS.454.3816C}
{Childress}, M.~J., {Hillier}, D.~J., {Seitenzahl}, I., {et~al.} 2015, \mnras,
  454, 3816

\bibitem[{{Chomiuk} {et~al.}(2016){Chomiuk}, {Soderberg}, {Chevalier},
  {Bruzewski}, {Foley}, {Parrent}, {Strader}, {Badenes}, {Fransson}, {Kamble},
  {Margutti}, {Rupen}, \& {Simon}}]{2016ApJ...821..119C}
{Chomiuk}, L., {Soderberg}, A.~M., {Chevalier}, R.~A., {et~al.} 2016, \apj,
  821, 119

\bibitem[{{Chotard} {et~al.}(2011){Chotard}, {Gangler}, {Aldering},
  {Antilogus}, {Aragon}, {Bailey}, {Baltay}, {Bongard}, {Buton}, {Canto},
  {Childress}, {Copin}, {Fakhouri}, {Hsiao}, {Kerschhaggl}, {Kowalski},
  {Loken}, {Nugent}, {Paech}, {Pain}, {Pecontal}, {Pereira}, {Perlmutter},
  {Rabinowitz}, {Runge}, {Scalzo}, {Smadja}, {Tao}, {Thomas}, {Weaver}, {Wu},
  \& {Nearby Supernova Factory}}]{Chotard2011}
{Chotard}, N., {Gangler}, E., {Aldering}, G., {et~al.} 2011, \aap, 529, L4

\bibitem[{{Conley} {et~al.}(2008){Conley}, {Sullivan}, {Hsiao}, {Guy},
  {Astier}, {Balam}, {Balland}, {Basa}, {Carlberg}, {Fouchez}, {Hardin},
  {Howell}, {Hook}, {Pain}, {Perrett}, {Pritchet}, \&
  {Regnault}}]{2008ApJ...681..482C}
{Conley}, A., {Sullivan}, M., {Hsiao}, E.~Y., {et~al.} 2008, \apj, 681, 482

\bibitem[{{Crotts}(2015)}]{2015ApJ...804L..37C}
{Crotts}, A.~P.~S. 2015, \apjl, 804, L37

\bibitem[{{Deustua} {et~al.}(2016){Deustua}, {Mack}, {Bowers}, {Baggett},
  {Bajaj}, {Dahlen}, {Durbin}, {Gosmeyer}, {Gunning}, {Hammer}, {Hartig},
  {Khandrika}, {MacKenty}, {Ryan}, {Sabbi}, \& {Sosey}}]{2016wfc..rept....3D}
{Deustua}, S.~E., {Mack}, J., {Bowers}, A.~S., {et~al.} 2016, {UVIS 2.0
  Chip-dependent Inverse Sensitivity Values}, Tech. rep.

\bibitem[{{Dimitriadis} {et~al.}(2017){Dimitriadis}, {Sullivan}, {Kerzendorf},
  {Ruiter}, {Seitenzahl}, {Taubenberger}, {Doran}, {Gal-Yam}, {Laher},
  {Maguire}, {Nugent}, {Ofek}, \& {Surace}}]{2017MNRAS.468.3798D}
{Dimitriadis}, G., {Sullivan}, M., {Kerzendorf}, W., {et~al.} 2017, \mnras,
  468, 3798

\bibitem[{{Drozdov} {et~al.}(2015){Drozdov}, {Leising}, {Milne}, {Pearcy},
  {Riess}, {Macri}, {Bryngelson}, \& {Garnavich}}]{2015ApJ...805...71D}
{Drozdov}, D., {Leising}, M.~D., {Milne}, P.~A., {et~al.} 2015, \apj, 805, 71

\bibitem[{{Folatelli} {et~al.}(2010){Folatelli}, {Phillips}, {Burns},
  {Contreras}, {Hamuy}, {Freedman}, {Persson}, {Stritzinger}, {Suntzeff},
  {Krisciunas}, {Boldt}, {Gonz{\'a}lez}, {Krzeminski}, {Morrell}, {Roth},
  {Salgado}, {Madore}, {Murphy}, {Wyatt}, {Li}, {Filippenko}, \&
  {Miller}}]{2010AJ....139..120F}
{Folatelli}, G., {Phillips}, M.~M., {Burns}, C.~R., {et~al.} 2010, \aj, 139,
  120

\bibitem[{{Foley} {et~al.}(2014){Foley}, {Fox}, {McCully}, {Phillips}, {Sand},
  {Zheng}, {Challis}, {Filippenko}, {Folatelli}, {Hillebrandt}, {Hsiao}, {Jha},
  {Kirshner}, {Kromer}, {Marion}, {Nelson}, {Pakmor}, {Pignata}, {R{\"o}pke},
  {Seitenzahl}, {Silverman}, {Skrutskie}, \&
  {Stritzinger}}]{2014MNRAS.443.2887F}
{Foley}, R.~J., {Fox}, O.~D., {McCully}, C., {et~al.} 2014, \mnras, 443, 2887

\bibitem[{{Fossey} {et~al.}(2014){Fossey}, {Cooke}, {Pollack}, {Wilde}, \&
  {Wright}}]{2014CBET.3792....1F}
{Fossey}, S.~J., {Cooke}, B., {Pollack}, G., {Wilde}, M., \& {Wright}, T. 2014,
  Central Bureau Electronic Telegrams, 3792

\bibitem[{{Fransson} \& {Jerkstrand}(2015)}]{2015ApJ...814L...2F}
{Fransson}, C., \& {Jerkstrand}, A. 2015, \apjl, 814, L2

\bibitem[{{Fransson} \& {Kozma}(1993)}]{1993ApJ...408L..25F}
{Fransson}, C., \& {Kozma}, C. 1993, \apjl, 408, L25

\bibitem[{{Graham} {et~al.}(2015){Graham}, {Nugent}, {Sullivan}, {Filippenko},
  {Cenko}, {Silverman}, {Clubb}, \& {Zheng}}]{2015MNRAS.454.1948G}
{Graham}, M.~L., {Nugent}, P.~E., {Sullivan}, M., {et~al.} 2015, \mnras, 454,
  1948

\bibitem[{{Graur} {et~al.}(2014){Graur}, {Maoz}, \&
  {Shara}}]{2014MNRAS.442L..28G}
{Graur}, O., {Maoz}, D., \& {Shara}, M.~M. 2014, \mnras, 442, L28

\bibitem[{{Graur} \& {Woods}(2018)}]{2018arXiv181104944G}
{Graur}, O., \& {Woods}, T.~E. 2018, ArXiv e-prints, arXiv:1811.04944

\bibitem[{{Graur} {et~al.}(2016){Graur}, {Zurek}, {Shara}, {Riess},
  {Seitenzahl}, \& {Rest}}]{2016ApJ...819...31G}
{Graur}, O., {Zurek}, D., {Shara}, M.~M., {et~al.} 2016, \apj, 819, 31

\bibitem[{{Graur} {et~al.}(2018{\natexlab{a}}){Graur}, {Zurek}, {Rest},
  {Seitenzahl}, {Shappee}, {Fisher}, {Guillochon}, {Shara}, \&
  {Riess}}]{2018ApJ...859...79G}
{Graur}, O., {Zurek}, D.~R., {Rest}, A., {et~al.} 2018{\natexlab{a}}, \apj,
  859, 79

\bibitem[{{Graur} {et~al.}(2018{\natexlab{b}}){Graur}, {Zurek}, {Cara}, {Rest},
  {Seitenzahl}, {Shappee}, {Shara}, \& {Riess}}]{2018ApJ...866...10G}
{Graur}, O., {Zurek}, D.~R., {Cara}, M., {et~al.} 2018{\natexlab{b}}, \apj,
  866, 10

\bibitem[{{Hack} {et~al.}(2012){Hack}, {Dencheva}, {Fruchter}, {Armstrong},
  {Avila}, {Baggett}, {Bray}, {Droettboom}, {Dulude}, {Gonzaga}, {Grogin},
  {Kozhurina-Platais}, {Lucas}, {Mack}, {MacKenty}, {Petro}, {Pirzkal},
  {Rajan}, {Smith}, {Sontag}, \& {Ubeda}}]{2012AAS...22013515H}
{Hack}, W.~J., {Dencheva}, N., {Fruchter}, A.~S., {et~al.} 2012, in American
  Astronomical Society Meeting Abstracts, Vol. 220, American Astronomical
  Society Meeting Abstracts \#220, 135.15

\bibitem[{{Horesh} {et~al.}(2012){Horesh}, {Kulkarni}, {Fox}, {Carpenter},
  {Kasliwal}, {Ofek}, {Quimby}, {Gal-Yam}, {Cenko}, {de Bruyn}, {Kamble},
  {Wijers}, {van der Horst}, {Kouveliotou}, {Podsiadlowski}, {Sullivan},
  {Maguire}, {Howell}, {Nugent}, {Gehrels}, {Law}, {Poznanski}, \&
  {Shara}}]{2012ApJ...746...21H}
{Horesh}, A., {Kulkarni}, S.~R., {Fox}, D.~B., {et~al.} 2012, \apj, 746, 21

\bibitem[{{Jacobson-Gal{\'a}n} {et~al.}(2018){Jacobson-Gal{\'a}n},
  {Dimitriadis}, {Foley}, \& {Kilpatrick}}]{2018ApJ...857...88J}
{Jacobson-Gal{\'a}n}, W.~V., {Dimitriadis}, G., {Foley}, R.~J., \&
  {Kilpatrick}, C.~D. 2018, \apj, 857, 88

\bibitem[{{Kelly} {et~al.}(2014){Kelly}, {Fox}, {Filippenko}, {Cenko}, {Prato},
  {Schaefer}, {Shen}, {Zheng}, {Graham}, \& {Tucker}}]{2014ApJ...790....3K}
{Kelly}, P.~L., {Fox}, O.~D., {Filippenko}, A.~V., {et~al.} 2014, \apj, 790, 3

\bibitem[{{Kerzendorf} {et~al.}(2017){Kerzendorf}, {McCully}, {Taubenberger},
  {Jerkstrand}, {Seitenzahl}, {Ruiter}, {Spyromilio}, {Long}, \&
  {Fransson}}]{2017MNRAS.472.2534K}
{Kerzendorf}, W.~E., {McCully}, C., {Taubenberger}, S., {et~al.} 2017, \mnras,
  472, 2534

\bibitem[{{Li} {et~al.}(2011){Li}, {Bloom}, {Podsiadlowski}, {Miller}, {Cenko},
  {Jha}, {Sullivan}, {Howell}, {Nugent}, {Butler}, {Ofek}, {Kasliwal},
  {Richards}, {Stockton}, {Shih}, {Bildsten}, {Shara}, {Bibby}, {Filippenko},
  {Ganeshalingam}, {Silverman}, {Kulkarni}, {Law}, {Poznanski}, {Quimby},
  {McCully}, {Patel}, {Maguire}, \& {Shen}}]{Li2011fe}
{Li}, W., {Bloom}, J.~S., {Podsiadlowski}, P., {et~al.} 2011, \nat, 480, 348

\bibitem[{{Maoz} \& {Graur}(2017)}]{2017ApJ...848...25M}
{Maoz}, D., \& {Graur}, O. 2017, \apj, 848, 25

\bibitem[{{Maoz} {et~al.}(2014){Maoz}, {Mannucci}, \&
  {Nelemans}}]{2014ARA&A..52..107M}
{Maoz}, D., {Mannucci}, F., \& {Nelemans}, G. 2014, \araa, 52, 107

\bibitem[{{Margutti} {et~al.}(2014){Margutti}, {Parrent}, {Kamble},
  {Soderberg}, {Foley}, {Milisavljevic}, {Drout}, \&
  {Kirshner}}]{2014ApJ...790...52M}
{Margutti}, R., {Parrent}, J., {Kamble}, A., {et~al.} 2014, \apj, 790, 52

\bibitem[{{Margutti} {et~al.}(2012){Margutti}, {Soderberg}, {Chomiuk},
  {Chevalier}, {Hurley}, {Milisavljevic}, {Foley}, {Hughes}, {Slane},
  {Fransson}, {Moe}, {Barthelmy}, {Boynton}, {Briggs}, {Connaughton}, {Costa},
  {Cummings}, {Del Monte}, {Enos}, {Fellows}, {Feroci}, {Fukazawa}, {Gehrels},
  {Goldsten}, {Golovin}, {Hanabata}, {Harshman}, {Krimm}, {Litvak},
  {Makishima}, {Marisaldi}, {Mitrofanov}, {Murakami}, {Ohno}, {Palmer},
  {Sanin}, {Starr}, {Svinkin}, {Takahashi}, {Tashiro}, {Terada}, \&
  {Yamaoka}}]{2012ApJ...751..134M}
{Margutti}, R., {Soderberg}, A.~M., {Chomiuk}, L., {et~al.} 2012, \apj, 751,
  134

\bibitem[{{Marion} {et~al.}(2015){Marion}, {Sand}, {Hsiao}, {Banerjee},
  {Valenti}, {Stritzinger}, {Vink{\'o}}, {Joshi}, {Venkataraman}, {Ashok},
  {Amanullah}, {Binzel}, {Bochanski}, {Bryngelson}, {Burns}, {Drozdov},
  {Fieber-Beyer}, {Graham}, {Howell}, {Johansson}, {Kirshner}, {Milne},
  {Parrent}, {Silverman}, {Vervack}, \& {Wheeler}}]{2015ApJ...798...39M}
{Marion}, G.~H., {Sand}, D.~J., {Hsiao}, E.~Y., {et~al.} 2015, \apj, 798, 39

\bibitem[{{Milne} {et~al.}(2001){Milne}, {The}, \&
  {Leising}}]{2001ApJ...559.1019M}
{Milne}, P.~A., {The}, L.-S., \& {Leising}, M.~D. 2001, \apj, 559, 1019

\bibitem[{{Munari} {et~al.}(2013){Munari}, {Henden}, {Belligoli}, {Castellani},
  {Cherini}, {Righetti}, \& {Vagnozzi}}]{2013NewA...20...30M}
{Munari}, U., {Henden}, A., {Belligoli}, R., {et~al.} 2013, \na, 20, 30

\bibitem[{{Penney} \& {Hoeflich}(2014)}]{2014ApJ...795...84P}
{Penney}, R., \& {Hoeflich}, P. 2014, \apj, 795, 84

\bibitem[{{Phillips}(1993)}]{1993ApJ...413L.105P}
{Phillips}, M.~M. 1993, \apjl, 413, L105

\bibitem[{{Quinn} {et~al.}(2006){Quinn}, {Garnavich}, {Li}, {Panagia}, {Riess},
  {Schmidt}, \& {Della Valle}}]{2006ApJ...652..512Q}
{Quinn}, J.~L., {Garnavich}, P.~M., {Li}, W., {et~al.} 2006, \apj, 652, 512

\bibitem[{{Schlafly} \& {Finkbeiner}(2011)}]{2011ApJ...737..103S}
{Schlafly}, E.~F., \& {Finkbeiner}, D.~P. 2011, \apj, 737, 103

\bibitem[{{Schmidt} {et~al.}(1994){Schmidt}, {Kirshner}, {Leibundgut}, {Wells},
  {Porter}, {Ruiz-Lapuente}, {Challis}, \& {Filippenko}}]{1994ApJ...434L..19S}
{Schmidt}, B.~P., {Kirshner}, R.~P., {Leibundgut}, B., {et~al.} 1994, \apjl,
  434, L19

\bibitem[{{Scolnic} {et~al.}(2014){Scolnic}, {Riess}, {Foley}, {Rest},
  {Rodney}, {Brout}, \& {Jones}}]{2014ApJ...780...37S}
{Scolnic}, D.~M., {Riess}, A.~G., {Foley}, R.~J., {et~al.} 2014, \apj, 780, 37

\bibitem[{{Seitenzahl} {et~al.}(2009){Seitenzahl}, {Taubenberger}, \&
  {Sim}}]{2009MNRAS.400..531S}
{Seitenzahl}, I.~R., {Taubenberger}, S., \& {Sim}, S.~A. 2009, \mnras, 400, 531

\bibitem[{{Shappee} {et~al.}(2017){Shappee}, {Stanek}, {Kochanek}, \&
  {Garnavich}}]{2017ApJ...841...48S}
{Shappee}, B.~J., {Stanek}, K.~Z., {Kochanek}, C.~S., \& {Garnavich}, P.~M.
  2017, \apj, 841, 48

\bibitem[{{Sparks} {et~al.}(1999){Sparks}, {Macchetto}, {Panagia}, {Boffi},
  {Branch}, {Hazen}, \& {Della Valle}}]{1999ApJ...523..585S}
{Sparks}, W.~B., {Macchetto}, F., {Panagia}, N., {et~al.} 1999, \apj, 523, 585

\bibitem[{{Taubenberger} {et~al.}(2015){Taubenberger}, {Elias-Rosa},
  {Kerzendorf}, {Hachinger}, {Spyromilio}, {Fransson}, {Kromer}, {Ruiter},
  {Seitenzahl}, {Benetti}, {Cappellaro}, {Pastorello}, {Turatto}, \&
  {Marchetti}}]{2015MNRAS.448L..48T}
{Taubenberger}, S., {Elias-Rosa}, N., {Kerzendorf}, W.~E., {et~al.} 2015,
  \mnras, 448, L48

\bibitem[{{Tody}(1986)}]{1986SPIE..627..733T}
{Tody}, D. 1986, in Society of Photo-Optical Instrumentation Engineers (SPIE)
  Conference Series, Vol. 627, Society of Photo-Optical Instrumentation
  Engineers (SPIE) Conference Series, ed. {D.~L.~Crawford}, 733--+

\bibitem[{{Tody}(1993)}]{1993ASPC...52..173T}
{Tody}, D. 1993, in Astronomical Society of the Pacific Conference Series,
  Vol.~52, Astronomical Data Analysis Software and Systems II, ed. R.~J.
  {Hanisch}, R.~J.~V. {Brissenden}, \& J.~{Barnes}, 173

\bibitem[{{Wang} {et~al.}(2018){Wang}, {Chen}, {Wang}, {Hu}, {Xi}, {Yang}, \&
  {Li}}]{2018arXiv181011936W}
{Wang}, X., {Chen}, J., {Wang}, L., {et~al.} 2018, ArXiv e-prints,
  arXiv:1810.11936

\bibitem[{{Wang} {et~al.}(2008){Wang}, {Li}, {Filippenko}, {Foley}, {Smith}, \&
  {Wang}}]{2008ApJ...677.1060W}
{Wang}, X., {Li}, W., {Filippenko}, A.~V., {et~al.} 2008, \apj, 677, 1060

\bibitem[{{Whelan} \& {Iben}(1973)}]{Whelan1973}
{Whelan}, J., \& {Iben}, Jr., I. 1973, \apj, 186, 1007

\bibitem[{{Yang} {et~al.}(2017){Yang}, {Wang}, {Baade}, {Brown}, {Cracraft},
  {H{\"o}flich}, {Maund}, {Patat}, {Sparks}, {Spyromilio}, {Stevance}, {Wang},
  \& {Wheeler}}]{2017ApJ...834...60Y}
{Yang}, Y., {Wang}, L., {Baade}, D., {et~al.} 2017, \apj, 834, 60

\bibitem[{{Yang} {et~al.}(2018){Yang}, {Wang}, {Baade}, {Brown}, {Cikota},
  {Cracraft}, {H{\"o}flich}, {Maund}, {Patat}, {Sparks}, {Spyromilio},
  {Stevance}, {Wang}, \& {Wheeler}}]{2018ApJ...852...89Y}
---. 2018, \apj, 852, 89

\end{thebibliography}
\end{document}